\documentclass{aipproc}
\layoutstyle{6x9}

\usepackage[utf8]{inputenc}
\usepackage[T1]{fontenc}
\usepackage[english]{babel}
\usepackage{amsfonts}
\usepackage{amssymb}
\usepackage{amsmath}
\usepackage{booktabs}
\usepackage{graphicx}
\usepackage{xcolor}
\usepackage{siunitx}
\usepackage{textgreek}
\usepackage{todonotes}
\usepackage{svg}
\usepackage{multirow}

\newcommand{\kr}{\textsuperscript{83m}Kr}
\newcommand{\rb}{\textsuperscript{83}Rb}

\newcommand{\krline}[3]{#1\textsubscript{#2}-#3{}}
\newcommand{\kline}[1]{\krline{K}{}{32}{}}
\newcommand{\lline}[1]{\krline{L}{}{32}{}}
\newcommand{\mline}[1]{\krline{M}{}{32}{}}

\begin{document}
\title{Characterization of the Detector Response to Electrons of Silicon Drift Detectors for the TRISTAN Project}

\author{M.~Lebert$ ^{1,2} $, T.~Brunst$^{1,2}$, T.~Houdy$^{1,2}$, S.~Mertens$^{1,2}$, and D.~Siegmann$^{1,2}$}{address={
$^1$ Max-Planck-Institut f\"ur Physik, F\"ohringer Ring 6, D-80805 M\"unchen, Germany\\
$^2$ Physik-Department, Technische Universit\"at M\"unchen, D-85747 Garching, Germany}}

%\author{Name Surname$^1$, Second Author$^2$, and A.~N. Other$^{1,2}$}{address={
%$^1$ Department of Physics, University in Paris, 123~45~Paris, France \\
%$^2$ Laboratory of Engineering, Institute of Science, 543~21~Tokyo, Japan}}
 % Please select relevant topics from: https://ufn.ru/en/pacs/

\begin{abstract} % Single paragraph.
	Right-handed neutrinos are a natural extension of the Standard Model of particle physics. 
	Such particles would only interact through the mixing with the left-handed neutrinos, hence they are called sterile neutrinos. 
	If their mass were in the \si{\kilo\electronvolt}-range they would be Dark Matter candidates. 
	By investigating the electron spectrum of the tritium \textbeta{}-decay the parameter space with masses up to the endpoint of \SI{18.6}{\kilo\electronvolt} can be probed. 
	A sterile neutrino manifests as a kink-like structure in the spectrum. 
	To achieve this goal the TRISTAN project develops a new detector system for the KATRIN experiment that can search for these new particles using the silicon drift detector technology. 
	One major effect on the performance of the detectors is the so called dead layer. 
	Here, a new characterization method for the prototype detectors is presented using the \textsuperscript{83m}Kr decay conversion electrons.
	By tilting the detector its effective dead layer increases, which leads to different peak positions. 
	The difference of peak positions between two tilting angles is independent of source effects and is thus suitable for characterization.
	A dead layer is then extracted by comparing the measurements to Monte-Carlo simulations. 
	A dead layer in the order of \SI{50}{\nano\meter} was found.
\end{abstract}

\maketitle

\section{Introduction}
    In the Standard Model of particle physics the neutrinos are massless, neutral leptons. 
    Those are the only particles that appear solely with left-handed chirality.  
    The detection of neutrino oscillation by the SNO and Super-Kamiokande Collaborations \cite{SNO,SuperK} showed that neutrinos are in fact not massless. 
    Therefore, the Standard Model is incomplete. 
    One possible natural extension would be the introduction of right-handed neutrinos. 
    These would only interact through the mixing with the known neutrinos and are hence called sterile neutrinos. 

    Each sterile neutrino would introduce a new mass eigenstate which is mostly made of the sterile. 
    Thus, the mixing with the known neutrinos would be small.    
    Depending on the mass scale of the additional mass eigenstate the sterile neutrino could solve different open questions of particle physics. 
    If the mass is heavy ($ \mathcal{O}(\si{\mega\electronvolt})$) the small masses of the active neutrinos could be explained by the seesaw mechanism. 
    For lighter masses ($ \mathcal{O}(\si{\kilo\electronvolt}) $) these new particles would be a viable Dark Matter candidate \cite{white}.   
    
\begin{figure}
    	\centering
    	\includegraphics[width=0.7\linewidth]{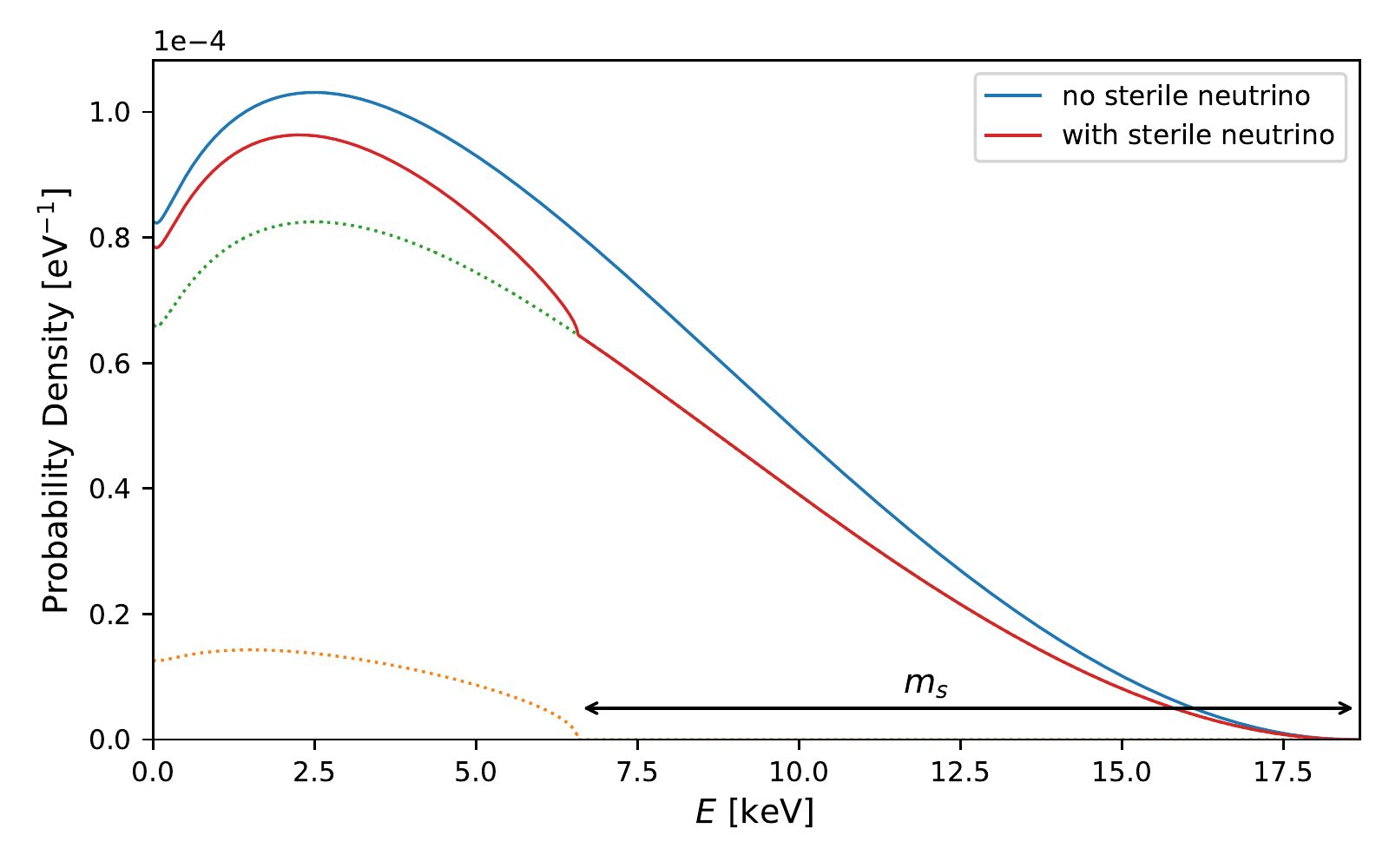}
    	\caption{Effect of a sterile neutrino mass on the \textbeta-spectrum. A kink appears at an energy below the endpoint of the \textbeta-spectrum corresponding the mass of the sterile neutrino. In this case a mass of \SI{12}{\keV} and a mixing of $ sin^{2}(\Theta) = \num{0.2} $ is shown. Taken from \cite{Mertens.2019}.}
    	\label{fig:sterilespectrum}
    \end{figure}

    To search for \si{\kilo\electronvolt}-sterile neutrinos the electron spectrum of the tritium \textbeta-decay can be investigated. 
    The neutrino reduces the maximum energy of the electron corresponding to the mass eigenstate that was created.  
    The \textbeta-spectrum is a superposition of all mass eigenstates that can be created in the decay. 
    Therefore, a heavy sterile mass eigenstate would manifest itself as a kink-like structure deep in the spectrum (fig.~\ref{fig:sterilespectrum}). 
    
    The amplitude of the kink and therefore the mixing is extremely small ($ sin^2(\Theta)~<~10^{-3} $). 
    To find such a small structure up to \num{e18} events are needed and thus a source with high luminosity is required \cite{Mertens.2019}. 
    This makes the Karlsruhe Tritium Neutrino (KATRIN) experiment predestined for such a task. 
    The windowless gaseous tritium source of KATRIN can provide up to \SI{e11}{decays\per\second} \cite{KATRINDesign}. 
    Currently the detector used for the neutrino mass measurements cannot handle the high rates needed. 
    Therefore, the Tritium Investigation on Sterile (A) Neutrinos (TRISTAN) project develops a new detector system using the Silicon Drift Detector (SDD) technology. 
    Rates of up to \SI{e8}{counts\per\second} on the detector are a compromise between the data collection time and the necessary development effort of the detector \cite{Mertens.2019}. 
    To handle these rates the system will distribute it across roughly \num{3500} pixels. 
    Also, to not wash out the kink the energy resolution must not exceed \SI{300}{\electronvolt} at \SI{30}{\kilo\electronvolt}. 
    Another important requirement is that the pixels need thin entrance windows. 
    Electrons are charged particles and have therefore a high interaction rate in matter through which they lose energy. 
    In the entrance window of the detector the energy collection of the detector is reduced and therefore deposited energy in this region is at least partly lost. 
    This effect would shift a peak of monoenergetic electrons to lower energies and create a low energy tail. 
    Therefore, it also effects the overall \textbeta-spectrum, hence it must be minimized and understood very well.

\begin{figure}
        \centering
        \includegraphics[width=0.5\textwidth]{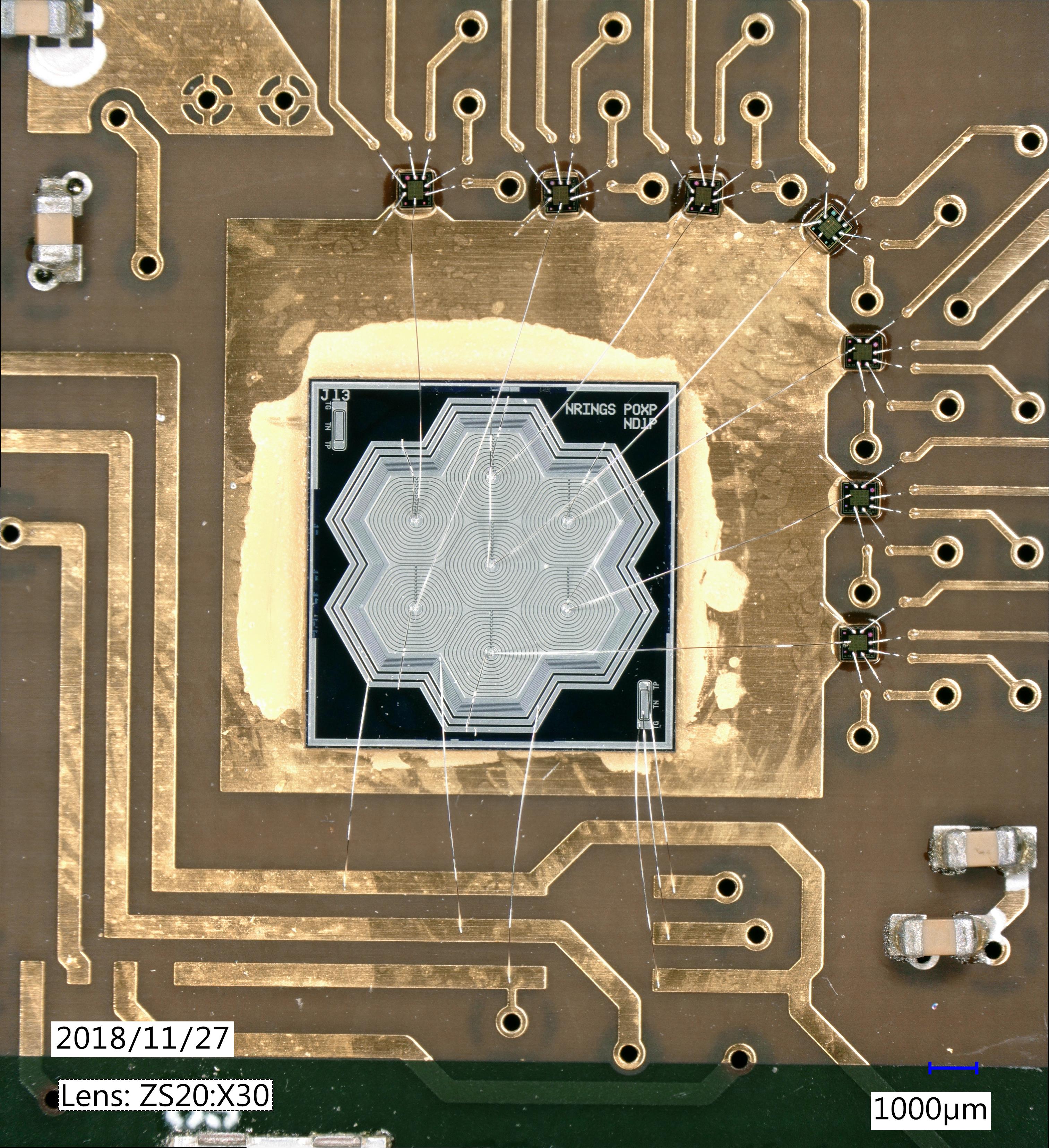}
        \caption{Back of a seven pixel prototype detector of the TRISTAN project.}
        \label{fig:detector}
    \end{figure}

    The aim of this work is to investigate a characterization method of the entrance window of the TRISTAN prototype detectors. 
    Different ion implantation techniques were implemented for the different prototype detectors. 
    All prototype detectors have seven, hexagonal shaped pixels, see fig.~\ref{fig:detector}. 
    A dead layer model is assumed to describe the entrance window of the detectors. 
    In this model a certain area in the beginning of the detector is completely insensitive, followed by a complete energy collection for the rest of the detector. 

\section{\textsuperscript{83}R\MakeLowercase{b}\textbackslash \textsuperscript{83\MakeLowercase{m}}K\MakeLowercase{r} sources}
\label{sec:Kr}
    The monoenergetic electron source used in this characterization is \kr{}. 
    This isomeric krypton is created through the decay of \rb{} via electron capture with a half-life of \SI{86.6}{\day}. 
    The metastable krypton state has an energy of about \SI{41.6}{\keV} and a half-half life of \SI{1.86}{\hour} \cite{McCutchan.2015}.
    This decays into the ground state in two steps. 
    The first decay releases \SI{32.2}{\keV} and the second one \SI{9.4}{\keV} \cite{McCutchan.2015}. 
    Conversion electrons can be created in both transitions but only from shells which binding energy is lower than the provided transition energy.
    This also leads to the emission of X-rays.  
    If no electron is emitted a \textgamma{}-photon is.

    For the investigations here only the X-ray peaks of K\textsubscript{\textalpha{}} and K\textsubscript{\textbeta{}} and the conversion electrons of the K-, L-, M-, and N-shell from the \SI{32.2}{\keV} transition are of interest. 
    The peaks are marked in fig.~\ref{fig:SourceComp}. 
    Conversion electrons from the \SI{9.4}{\keV} transition are in a low energy continuum and are therefore unsuited for characterization. 
    Energetic differences within one shell are too small to be distinguished with the TRISTAN SDDs. 
    Thus, the L-lines form one peak and the M- and N-lines form another one. 
    
    The calibration of the detector was done using the photon lines of \textsuperscript{241}Am because photons are hardly altered by the dead layer. 
    An advantage of \kr{} is the occurrence of photon and electron lines which makes an in-situ test of the calibration possible. 
    
    In the setup of the calibration no gaseous source could be used. 
    For this reason rubidium is vacuum evaporated onto a backing of Highly Oriented Pyrolytic Graphite (HOPG) or rigid graphite  by the Nuclear Physics Institute \v{R}e\v{z}/Prague.
    Thereby roughly a mono-layer \rb{} is placed onto the graphite \cite{Venos.2010}. 
    Energy losses of electrons traveling through this layer are negligible. 
    This makes this kind of source acceptable for this characterization.

\section{Measurements}
    
\begin{figure}
    	\centering
    	\includegraphics[width=0.8\linewidth]{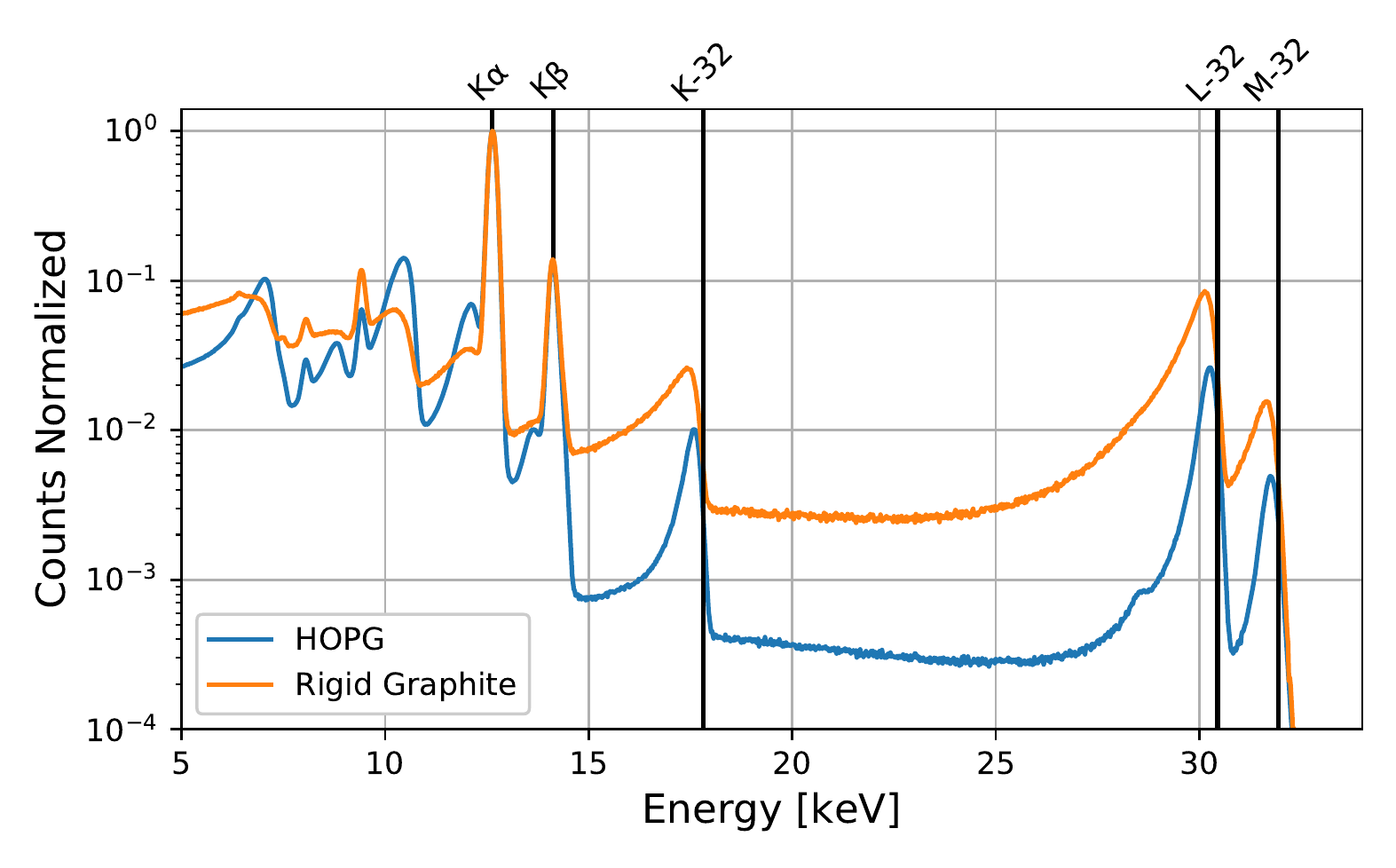}
    	\caption{Comparison of the spectra obtained from the decay of \kr{} in the HOPG and the rigid graphite source. It is visible that rigid graphite alters the electrons significantly. }
    	\label{fig:SourceComp}
    \end{figure}

    The sources are placed approximately \SI{1}{\centi\meter} below the detector in a vacuum chamber. 
    The resulting spectra obtained from the HOPG and rigid graphite sources are shown in fig.~\ref{fig:SourceComp}. 
    The X-ray peaks are Gaussian with only a small asymmetry to low energies. 
    Electron peaks on the other hand have a prominent low energy tail and are shifted to lower energies compared to the theoretical values.

    Figure~\ref{fig:SourceComp} also shows that the backing of the source has an influence on the conversion electrons. 
    The electron peaks of the rigid graphite source have much more pronounced low energy tails and are at slightly lower energies than the peaks from the HOPG source. 
    The electron background in the region from \SIrange[range-units=single]{18}{30}{\keV}, is again higher for the rigid graphite. 
    HOPG has lower peak amplitudes which comes from the different retention factors of the sources. 
    The retention factor states the probability that a bound \rb{} decays into the noble gas \kr{} which then stays in the position of the rubidium. 
    The noble gas is much weaker bound and can therefore leave the source before decaying. 
    
	This comparison shows that the source can potentially disturb or alter the characterization. 
	For this reason a method is needed that is independent of any source effects.

	The electron peak position $\Bar{E}_i(\theta)$ depends on the tilting angle $\theta$ between the detector and the source. 
	It can be described by:
	\begin{equation}
	\Bar{E}_i(\theta) = \bar{E}^{\mathrm{th}}_i - \Delta DL^i(\theta) - \delta_{\Phi} - \delta_{\mathrm{source}}.
	\label{eq:PeakPos}
	\end{equation}
	Here, the theoretical peak position is given by $\bar{E}^{\mathrm{th}}_i$, the potential applied to the entrance window side of the detector to deplete it is $\delta_\Phi$, and $\delta_{\mathrm{source}}$ are source effects. 
	$\Delta DL^i(\theta)$ is the lost energy of the electrons within the dead layer of the detector for a given peak $i$.
	It increases for higher tilting angles $\theta$ because the effective dead layer also increases and thus the energy loss.
	Therefore, an energy difference $\Delta$ occurs if two measurements at different angles are compared. 
	This shift is independent of any external effects and only depends on the dead layer:
	\begin{equation}
	\Delta = \Bar{E}_i(\theta) - \Bar{E}_i(\theta') = \Delta DL^i(\theta') - \Delta DL^i(\theta).
	\end{equation}
	
	The measured shifts for the conversion electron peaks from the \SI{32.2}{\keV} transition and the HOPG source are given in table~\ref{tab:shifts}. 
	A decreasing energy shift for higher electron energies appears due to the longer mean free path. 
	Based on Monte-Carlo simulations with the KATRIN simulation package KESS~\cite{KESS}, we can relate the measured energy shift to a step-like dead layer of about \SI{50}{\nano\meter}, see table~\ref{tab:shifts}.

	\begin{table}
		\centering

			\centering
			\caption{Dead layer shifts $\Delta^i_{DL}$ for one prototype detector and the corresponding dead layer thicknesses resulting from the comparison between measurement and simulation.}
			\label{tab:shifts}%
%			\centering
			\sisetup{table-number-alignment=center}
			\begin{tabular}{cScSS}
				
				\toprule
				Peak &{Energy}		& Shift &  {Shift} & {Dead Layer}\\
					& {[\si{\keV}]} & 		& {[\si{\electronvolt}]} & {[\si{nm}]} \\
				\midrule
				\midrule
				\kline{}		&\num{17.8}  	& $\Delta^K_{DL}$	 	& \num{45 \pm 5} 				&\num{50 \pm 6}  \\
				\lline{}		&\num{30.4}  	& $\Delta^L_{DL}$	 	& \num{32 \pm 5} 				&\num{55 \pm 6}   \\
				\mline{}		&\num{31.9}		& $\Delta^M_{DL}$		& \num{29 \pm 6} 	 				&\num{55 \pm 9}   \\
				\bottomrule
			\end{tabular}

	\end{table}% 

\section{Conclusion}
	
	It can be seen that the determined dead layers for each detector are, as expected, constant for the different energies.  
	Previous measurements with an electron microscope at \SI{14}{\keV} had energy shifts of \SI{45+-1}{\electronvolt} and a dead layer of \SI{46+-6}{\nano\meter} \cite{Siegmann.2019}. 
	These values are compatible with the krypton measurements. 
	This shows that the tilt-method can be used with an electron microscope or conversion electrons from \kr{} as source which can then cross-check each other. 
	Therefore, this method is suitable for the characterization of the TRISTAN detectors. 
	Investigations of a more detailed depth-dependent charge-collection efficiency model and comparisons to Geant4~\cite{Geant4} simulations are currently ongoing.

%\begin{theacknowledgments}
%This work was supported by the\dots \todo{Do I need acknowledgments?}
%\end{theacknowledgments}

\bibliographystyle{aipproc}

\begin{thebibliography}{99} % 9 for less than 10 references, else 99 for less than 100 references, etc.
	\bibitem{SNO}
        Q.~R.~Ahmad \emph{et al.} (SNO Collaboration), \emph{Phys. Rev. Lett.}, \textbf{89}, 011301 (2002)
    \bibitem{SuperK}
        Y.~Fukuda \emph{et al.} (Super-Kamiokande Collaboration), \emph{Phys. Rev. Lett.}, \textbf{81}, 1562 (1998)
    \bibitem{white}
    	R.~Adhikari \emph{et al.}, \emph{J. Cosmol. Astropart. Phys.}, \textbf{01}, 025 (2017)
	\bibitem{Mertens.2019}
		S.~Mertens \emph{et al.}, \emph{J. Phys. G: Nucl. Part. Phys.}, \textbf{46}, 065203 (2019)
	\bibitem{KATRINDesign}
        KATRIN Collaboration, KATRIN design report 2004 (2005)
	\bibitem{Venos.2018}
	    D.~Vénos \emph{et al.}, \emph{JINST}, \textbf{13}, T02012 (2018)
     \bibitem{McCutchan.2015}
        E.~A.~McCutchan, \emph{NUCL DATA SHEETS}, \textbf{125}, 201-394 (2018)
    \bibitem{Venos.2010}
        D.~Vénos \emph{et al.}, \emph{Meas Tech} \textbf{53}, 573 (2010)
    \bibitem{KESS}
	    P.~Renschler, "KESS - A new Monte Carlo simulation code for low-energy electron interactions in silicon detectors", Phd. diss, Karlsruher Institut f\"ur Technologie (2011)
    \bibitem{Siegmann.2019}
    	D.Siegmann, "Investigation of the Detector Response to Electrons of the TRISTAN Prototype Detectors", Master's thesis, Technische Universit\"at M\"unchen (2019)
    \bibitem{Geant4}
    	J.~Allison \emph{et al.}, \emph{Nucl. Instrum. Methods Phys. Res. A}, \textbf{835}, 186-225 (2016)


    
  

\end{thebibliography}

\newpage
\end{document}